\documentclass[aps,pra,twocolumn,groupedaddress,showpacs]{revtex4-1}
\usepackage{graphicx}
\usepackage{amsmath}
\usepackage{amsfonts}
\usepackage{amssymb}
\usepackage[normalem]{ulem}
\usepackage{xcolor}
\usepackage{soul}
\graphicspath{{./Figures/}}
\usepackage{float}

\def\ket#1{{\left\vert#1\right\rangle}}
\def\abs#1{{\left|#1\right|}}

\def\tFWHM{\tau_\text{FWHM}}
\def\tctrl{\tau^\text{ctrl}_\text{FWHM}}
\def\Deltau{\Delta\tau^\text{ctrl}}

\definecolor{darkgreen}{RGB}{34,139,34}

\begin{document}

\title{Lineshape Optimization in Inhomogeneous $\Lambda$-type Quantum Memory}
\author{Kai Shinbrough$^{1,2}$$^\dagger$}
\email{kais@illinois.edu}
\author{Donny R. Pearson Jr.$^{1,2}$}
\thanks{These authors contributed equally to this work}
\author{Virginia O. Lorenz$^{1,2}$}
\author{Elizabeth A. Goldschmidt$^{1,2}$}

\address{$^1$Department of Physics, University of Illinois at Urbana-Champaign, 1110 West Green Street, Urbana, IL 61801, USA\\
$^2$Illinois Quantum Information Science and Technology (IQUIST) Center, University of Illinois at Urbana-Champaign, 1101 West Springfield Avenue, Urbana, IL 61801, USA}

\date{\today}

\begin{abstract}
Photonic quantum memory is a crucial elementary operation in photonic quantum information processing. While many physically distinct memory protocols and hardware implementations have been 
applied to this task, the development of a quantum memory 
performant in all relevant metrics simultaneously (e.g., efficiency, bandwidth, lifetime, etc.) is still an open challenge. In this work, we focus on inhomogeneously broadened ensembles of $\Lambda$-type quantum emitters, which have long coherence lifetimes and broad bandwidth compatibility, but tend to exhibit low efficiency, in part due to technical constraints on medium growth and preparation, and in part due to inefficient use of a key resource in these systems: the inhomogeneously broadened excited state lineshape. We investigate the properties of electromagnetically induced transparency (EIT) for a survey of inhomogeneous lineshapes that are straightforward to realize experimentally, and optimize the memory efficiency for each lineshape over a large range of experimental parameters. We compare the optimal EIT efficiency to the well-known atomic frequency comb (AFC) protocol, which also relies on spectral shaping of the inhomogeneous broadening, and observe that with sufficient control field power the optimized lineshapes allow more efficient storage. Finally, we optimize over the inhomogeneous lineshape in a protocol agnostic fashion by numerically constructing the linear integral kernel describing the memory interaction and using a singular value decomposition and interpolation procedure to ensure optimality of the resulting lineshape. 
\end{abstract}

\maketitle

\section{Introduction}

The search for high-performing photonic quantum memories is motivated by memory-enhanced applications in quantum communication \cite{sangouard2011quantum,duan2001long}, computation \cite{knill2001scheme,raussendorf2001one}, and metrology \cite{khabiboulline2019optical,gottesman2012longer}. A quantum memory useful for these applications ought to transduce incoming photonic qubits into stationary excitations, protect these 
excitations from decay and dephasing for an appreciable amount of time, and transduce back to the photonic domain in a quantum-information-preserving manner with high efficiency, low-noise, and high throughput. A wide variety of physical platforms and theoretical protocols have been developed for this task in recent years (for reviews see, e.g., Refs.~\cite{shinbrough2023broadband,lvovsky2009optical,simon2010quantum,ma2017optical,novikova2012electromagnetically,heshami2016quantum}). Inhomogeneously broadened ensembles of $\Lambda$-type quantum emitters, in particular rare-earth-ions (REIs) doped in solids, have gathered significant interest due to their long coherence lifetimes \cite{businger2022non,ranvcic2018coherence,ma2021one}, broad bandwidth compatibility \cite{askarani2019storage,saglamyurek2016multiplexed}, low-noise \cite{zhou2015quantum}, and ability to be fabricated in chip-scale waveguide structures \cite{liu2020demand, dutta2023atomic}.

Most optimization of 
quantum light storage protocols in these systems to date relies on the temporal/spectral shaping and timing of the optical control field mediating the memory interaction. This amounts to optimization of the dressed ensemble of three-level $\Lambda$-type emitters, where the signal field to be stored is interacting with the combined atom--control-field system. Another domain of optimization is accessible in some systems (like REI systems), which amounts to optimization of the bare atomic system before it is dressed by the control field. This extra degree of freedom allows for regimes that cannot be achieved by optimizing the control field alone. To date, most work investigating spectral manipulation of the bare atomic ensemble have not attempted to find an optimal scheme, but have rather focused on a few simple cases \cite{hedges2010efficient,lauro2009slow}. 
Here we build on previous work optimizing the control field parameters for $\Lambda$-type memory across different memory protocols 
\cite{shinbrough2021optimization,shinbrough2023broadband} 
to additionally optimize the bare inhomogeneous profile of the ensemble.

\begin{figure*}[t]
	\centering
	\includegraphics[width=1.5\columnwidth]{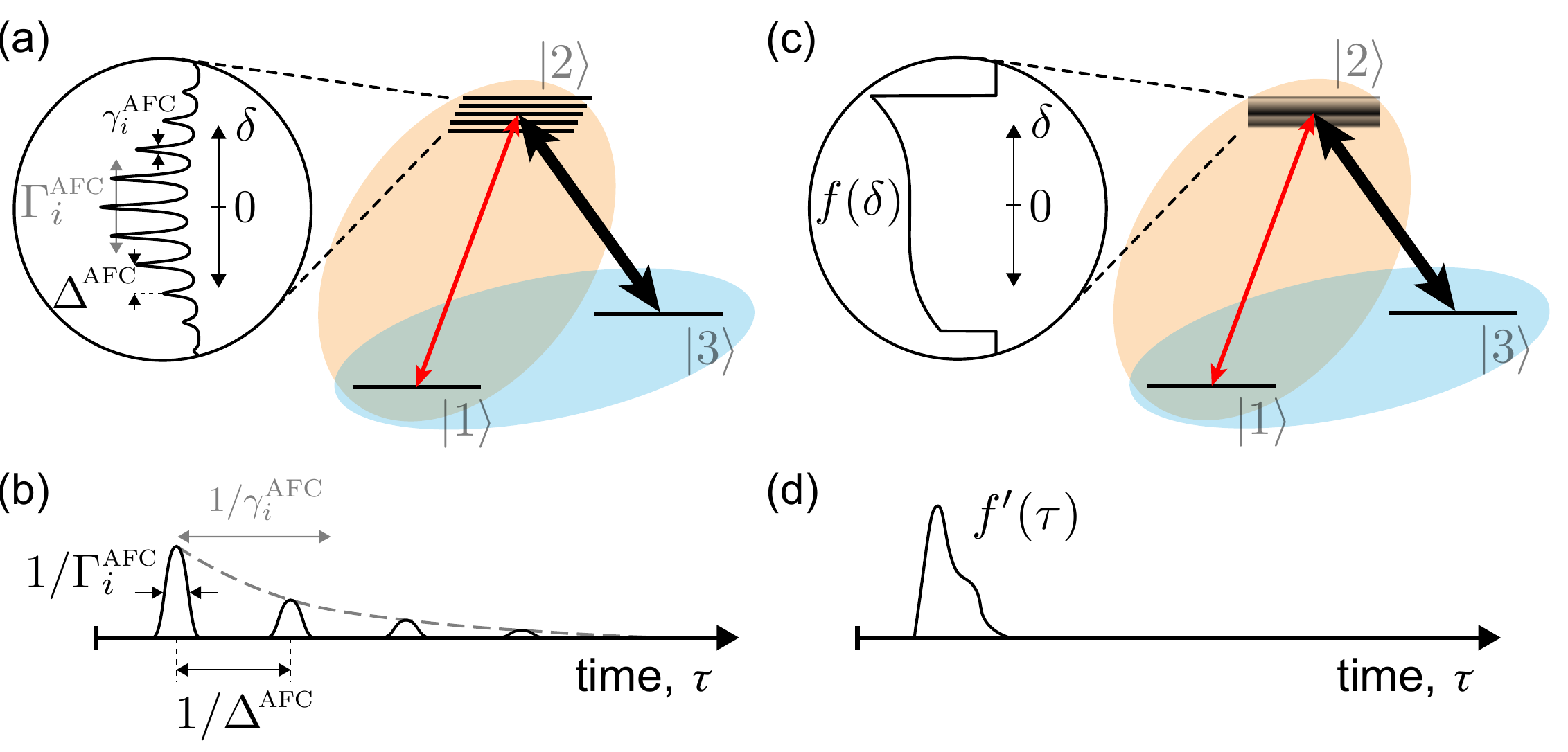}
	\caption{Schematic representations of (a) the $\Lambda$-type atomic frequency comb (AFC) protocol, (b) the resulting time trace of reemitted photons, (c) the $\Lambda$-type quantum memory protocol involving an arbitrarily shaped inhomogeneous profile, and (d) the resulting arbitrary time trace.}
	\label{Fig_AFC_v_arb}
\end{figure*}

In these inhomogeneous systems, 
a common memory protocol is the so-called atomic frequency comb (AFC) protocol \cite{afzelius2009multimode}, implemented in either a two- or three-level system (often referred to as optical and spin wave storage, respectively). In a three-level $\Lambda$-type system, shown in Fig.~\ref{Fig_AFC_v_arb}(a), the protocol proceeds as follows: First, a series of spectral holes are burned in the ground-to-excited state transition by shelving atoms resonant with the burn frequencies in a long-lived auxiliary state. These spectral holes are equidistant in frequency separation, $\Delta^\text{AFC}$, and therefore form a frequency comb in the ground-to-excited state transition absorption spectrum. The linewidth of each resulting tine in the frequency comb is denoted $\gamma_i^\text{AFC}$, and the bandwidth of the envelope of the absorption spectrum is labeled $\Gamma_i^\text{AFC}$. 
A single-photon-level quantum state [the signal field; thin, red arrow in Fig.~\ref{Fig_AFC_v_arb}(a)] is sent into the ensemble resonant with the center of the ground-to-excited state AFC. The absorption of this optical pulse results in a frequency-multimode atomic polarization [orange oval in Fig.~\ref{Fig_AFC_v_arb}(a)], which 
rephases after a time proportional to $1/\Delta^\text{AFC}$. 
Before rephasing, a $\pi$-pulse control field resonant with the excited-to-storage state transition transduces the atomic polarization into a spin excitation [blue oval in Fig.~\ref{Fig_AFC_v_arb}(a)]. Finally, retrieval in this protocol relies on application of a second $\pi$-pulse control field, re-population of the atomic polarization, and subsequent rephasing and optical reemission. This reemission occurs in sequential pulses, shown in Fig.~\ref{Fig_AFC_v_arb}(b), parameterized by the inverse of the AFC linewidths and detunings. 
This AFC protocol can thus be considered an absorb-then-transfer (ATT) protocol with additional optimization over the inhomogeneous profile of the atomic ensemble. In this work, we aim to extend this inhomogeneous profile optimization across all broadband regimes and memory protocols.

The AFC protocol is well-suited to a number of different quantum information processing tasks, owing to its ease of implementation, reasonable efficiency, and multimode capacity. The bandwidth of AFC memories can be large, though typically this precludes spin-state storage and thus limits the available storage time. 
Without the use of optical resonators and spin-wave storage, AFC quantum memories are fundamentally limited in efficiency and lifetime \cite{afzelius2009multimode}, and when such enhancements are introduced, these memories tend to be practically limited in efficiency and bandwidth  \cite{shinbrough2023broadband}. 
Several schemes exist to mitigate these limitations, typically at the expense of other memory performance metrics \cite{afzelius2010impedance,moiseev2010efficient}, but here we focus on the principle reason for this limitation: that the AFC protocol makes inefficient use of the total available atom number 
in inhomogeneously broadened media. In the canonical AFC protocol, the effective optical depth $d_\text{eff}$ of the AFC is reduced from the optical depth of the unmodified inhomogeneous distribution, $d$, by a factor proportional to the AFC finesse, $\mathcal{F}$: $d_\text{eff} = (d/\mathcal{F})\sqrt{\pi/4\ln2}$. The inverse of the AFC finesse ($1/\mathcal{F}$) represents the fraction of atoms that remain in the AFC after spectral hole burning. Alternative protocols that use a larger fraction of the atom number and lineshape initially present in the inhomogeneous distribution have been explored in Refs.~\cite{gorshkov2007photon_3,vivoli2013high}, and, recently, electromagnetically induced transparency (EIT) has been explored experimentally and theoretically for a variety of inhomogeneous lineshapes in Ref.~\cite{fan2019electromagnetically}. These works taken together lead to the important questions of whether an optimal inhomogeneous lineshape exists for a given inhomogeneously broadened $\Lambda$-type quantum memory, and whether such a memory protocol is more resource-efficient than others. 

A schematic of the general principle behind this work is given in Fig.~\ref{Fig_AFC_v_arb}(c) and (d), where we consider the arbitrarily shaped inhomogeneous profile $f(\delta)$ and the corresponding time trace of the reemitted pulse, $f'(\tau)$, where $\delta$ corresponds to the frequency shift from the center of the unmodified inhomogeneous profile and $\tau$ to time measured in the co-moving frame. In Section \ref{surveySec} we consider a survey of several common inhomogeneous lineshapes, for which the resulting electromagnetically induced transparency visibility and bandwidth are known \cite{fan2019electromagnetically}, and we calculate the corresponding quantum memory efficiencies in the regime of intermediate optical depth, photon bandwidth, and inhomogeneous linewidth. 
Furthermore, in Sec.~\ref{surveySec} we compare the memory efficiencies achieved using a finesse-optimized AFC versus a Rectangular inhomogeneous profile using the EIT memory protocol, for experimentally feasible optical depths and control field powers. 
In Section \ref{optSec} we consider the general case of arbitrary shape-based optimization of the inhomogeneous profile.

\section{Survey of Inhomogeneous Lineshapes}\label{surveySec}

In this section we aim to compare several common inhomogeneous lineshapes and probe their suitability for $\Lambda$-type quantum memory via the EIT, ATS
, and ATT 
protocols. Detailed descriptions of each protocol can be found in many other works, including, e.g., Refs.~\cite{gorshkov2007photon_2,gorshkov2007photon_3,saglamyurek2018coherent,vivoli2013high,shinbrough2021optimization,shinbrough2023broadband}; here we aim simply to optimize $\Lambda$-type quantum memory efficiency in inhomogeneously broadened media in a protocol-agnostic fashion. 

\begin{figure}[t]
	\centering
	\includegraphics[width=0.9\columnwidth]{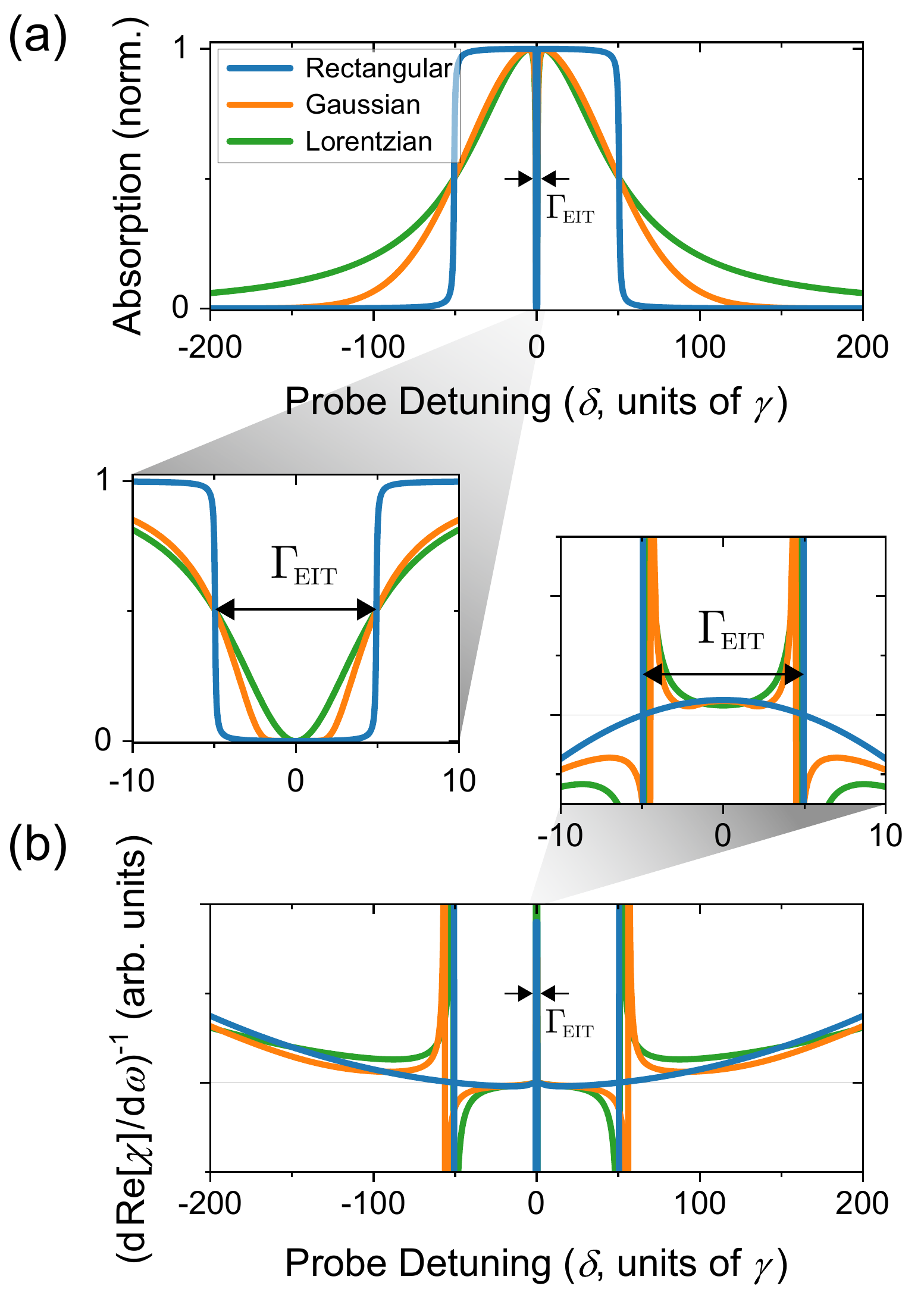}
	\caption{Electromagnetically induced transparency (EIT) in inhomogeneously broadened $\Lambda$-type ensembles with Rectangular, Gaussian, and Lorentzian lineshapes. (a) Normalized absorption and (b) inverse derivative of the real part of the linear susceptibility with respect to frequency, which is proportional to the group velocity, vs detuning from resonance. For all curves: $(\gamma_i,\Omega_c,\Delta) = (50\gamma,10\gamma,0\gamma)$ (insets: regions of transparency and slow light).}
	\label{Fig_abs_dreXdw}
\end{figure}

A simple calculation that demonstrates the importance of inhomogeneous lineshape in these systems involves the linear susceptibility in the presence of EIT. 
As shown in Ref.~\cite{fan2019electromagnetically}, this frequency-dependent susceptibility can be expressed in terms of the homogeneous susceptibility $\chi_h \propto 4\delta/\{\Omega_c^2+2i\delta[\gamma-2i(\delta-\delta')]\}$ and the inhomogeneous lineshape $f(\delta')$ as

\begin{equation}\label{inhomX}
    \chi(\delta) \propto \int d\delta' \, f(\delta')\chi_h(\delta,\delta'),
\end{equation}

\noindent where $\Omega_c$ is the Rabi frequency of the control field resonant with the $\ket{2}\leftrightarrow\ket{3}$ transition [bold, black arrow in Fig.~\ref{Fig_AFC_v_arb}(c); detuning from the center of the inhomogeneous distribution, $\Delta = 0$] and $\gamma$ is the homogeneous coherence decay rate of the excited state. 
The half width at half maximum of the excited state inhomogeneous distribution is $\gamma_i$, where $\gamma_i\gg\gamma$. For sake of simplicity, we assume the absence of inhomogeneity in the spin transition, $\ket{1}\leftrightarrow\ket{3}$, which would reduce the EIT visibility and provide a lower bound on the EIT FWHM linewidth, $\Gamma_\text{EIT}$, according to the analytic expressions of Ref.~\cite{fan2019electromagnetically}. The real part of this susceptibility 
determines the refractive index of the EIT medium given by $n = \sqrt{1+\text{Re}[\chi(\delta)]}$ and the resulting group velocity given by $v_g = c/[n + \omega(dn/d\omega)]$, where $\omega$ is the angular frequency of the signal field \cite{fleischhauer2005electromagnetically}. The imaginary part of the susceptibility determines the absorption lineshape, $\alpha \propto \text{Im}[\chi(\delta)]$ \cite{boyd2008nonlinear}. Here we consider the figure of merit $(d \text{Re}[\chi]/d\omega)^{-1}$ which captures the qualitative behavior of $v_g$ without loss of generality. In Fig.~\ref{Fig_abs_dreXdw}, we consider three common inhomogenous lineshapes, the Rectangular, Gaussian, and Lorentzian distributions, each with the EIT parameters $(\gamma_i,\Omega_c,\Delta) = (50\gamma,10\gamma,0\gamma)$. This simple calculation reveals the importance of the inhomogeneous lineshape: All three distributions lead to the same EIT FWHM linewidth and visibility [Fig.~\ref{Fig_abs_dreXdw}(a) and inset], but lead to significantly different EIT profiles and frequency-dependent group velocities [Fig.~\ref{Fig_abs_dreXdw}(b) and inset]. In particular, the Rectangular function possesses a deeper, flatter transparency window, as well as a significantly flatter group velocity profile, both of which could serve to improve $\Lambda$-type EIT quantum memory efficiency and bandwidth in inhomogeneously broadened media. This behavior is due to favorable constructive and destructive interference of homogeneous EIT profiles in Eq.~\eqref{inhomX}.

Rectangular inhomogeneous lineshapes possess a significant advantage in EIT profile relative to Gaussian or Lorentzian lineshapes of the same FWHM; a natural question arises as to whether this advantage persists for true $\Lambda$-type quantum memory operation, however. To answer this question, we consider the three-level inhomogeneous Maxwell-Bloch equations \cite{gorshkov2007photon_3,nunn2008quantum}:

\begin{align}
    \label{Aeq}\partial_z A &= -\sqrt{d} P_\text{tot}\\
    \label{Peq}\partial_\tau P_\delta &= -(\gamma-i\delta) P_\delta + \sqrt{d} \sqrt{f(\delta)}A - i\frac{\Omega_c}{2} B_\delta\\
    \label{Beq}\partial_\tau B_\delta &= -\gamma_B B_\delta -i\frac{\Omega_c^*}{2} P_\delta,
\end{align}

\noindent where each field---$A$ (the signal field), $P_\delta$ (the single-frequency-component atomic polarization), $B_\delta$ (the single-frequency-component spin excitation), and $P_\text{tot}$ (the frequency-multimode atomic polarization)---is implicitly $z$- and $\tau$-dependent, where $z$ is the one-dimensional spatial coordinate of the ensemble; we define $P_\text{tot}=\int d\delta \sqrt{f(\delta)} P_\delta$, and we assume a normalized lineshape such that $\int d\delta \, f(\delta) = 1$; $d$ is the optical depth at the peak of the absorption spectrum of the ensemble, and $\gamma_B$ is the coherence decay rate of the spin excitation. For signal field bandwidths much larger than the homogeneous excited state linewidth, $BW\gg\gamma$, the decay of $P_\delta$ in time is negligible and we can take $\gamma\rightarrow 0$ in line \eqref{Peq}. We assume incident signal field temporal envelopes that are Gaussian in shape, with full-width at half maximum (FWHM) duration $\tFWHM=2\pi\times 2\ln{2}/(\pi\, BW)$ [$A_\text{in}(\tau) = e^{-\tau^2/4\sigma^2}$, where $\sigma=\tFWHM/(2\sqrt{2\ln{2}})$]. A comparison of the memory efficiencies achieved with Gaussian and Lorentzian inhomogeneous lineshapes in the limiting case where $\tFWHM\gamma_i\rightarrow0$ and $d\rightarrow\infty$ but $d\tFWHM\gamma_i$ remains finite have been examined in Refs.~\cite{gorshkov2007photon_3,vivoli2013high}. Here, for the first time, we investigate the regime of intermediate $\tFWHM\gamma_i$ and $d$ for these lineshapes as well as the Rectangular lineshape examined above. 

We simulate the atomic dynamics in Eqs.~\eqref{Aeq}-\eqref{Beq} through Ralston's method for the $\tau$ derivatives and Euler's method for the $z$ derivatives. We find the runtime of these simulations is significantly reduced by normalizing the time dependence of Eqs.~\eqref{Aeq}-\eqref{Beq} by the inhomogeneous linewidth $\gamma_i$, and by assigning $P_\delta$ and $B_\delta$ to temporary vectors at each time step, recording only the full temporal evolution of $P_\text{tot}$ and $B_\text{tot}=\int d\delta \sqrt{f(\delta)} B_\delta$ in computer memory. In our simulations we consider 501 inhomogeneously distributed frequency classes spread uniformly in a range of $\delta~\sim~[-10\gamma_i,+10\gamma_i]$, which, in the worst case, ensures sampling of the inhomogeneous distribution out to below 1\% of its maximal value. At each set of memory parameters $\mathcal{M} = (d,\tFWHM\gamma_i)$, we follow the prescription of Refs.~\cite{shinbrough2021optimization,shinbrough2023variance} and optimize over the parameters 
of a Gaussian control field $\Omega(\tau) = \Omega_0\, e^{-[(\tau-\Deltau)/2\sigma^\text{ctrl}]^2}$, where $\Omega_0 = \theta/(2\sqrt{\pi}\sigma^\text{ctrl})$, $\theta = \int_{-\infty}^\infty d\tau\, \Omega(\tau)$ is the control field pulse area, $\Deltau$ is its temporal delay relative to the signal field, and $\tctrl=2\sqrt{2\ln{2}}\sigma^\text{ctrl}$ is the control field duration. Specifically, we optimize the parameter space vector $\mathcal{G} = \left(\theta, \Deltau, \tctrl\right)$ using a Nelder-Mead simplex method with respect to the memory efficiency,

\begin{equation}
    \eta = \frac{\int_{-\infty}^{\infty}\abs{A_\text{out}(\tau)}^2}{\int_{-\infty}^{\infty}\abs{A_\text{in}(\tau)}^2}.
\end{equation}

\noindent Here, we assume co-propagating signal and control fields in the storage operation, defining the forward $+z$ direction, and backward-propagating signal and control fields in the retrieval operation. Thus $A_\text{in}(\tau)$ refers to the temporal envelope of the forward-propagating input signal field, and $A_\text{out}(\tau)$ refers to the temporal envelope of the backward-propagating retrieved signal field.

\begin{figure}[t]
	\centering
	\includegraphics[width=\columnwidth]{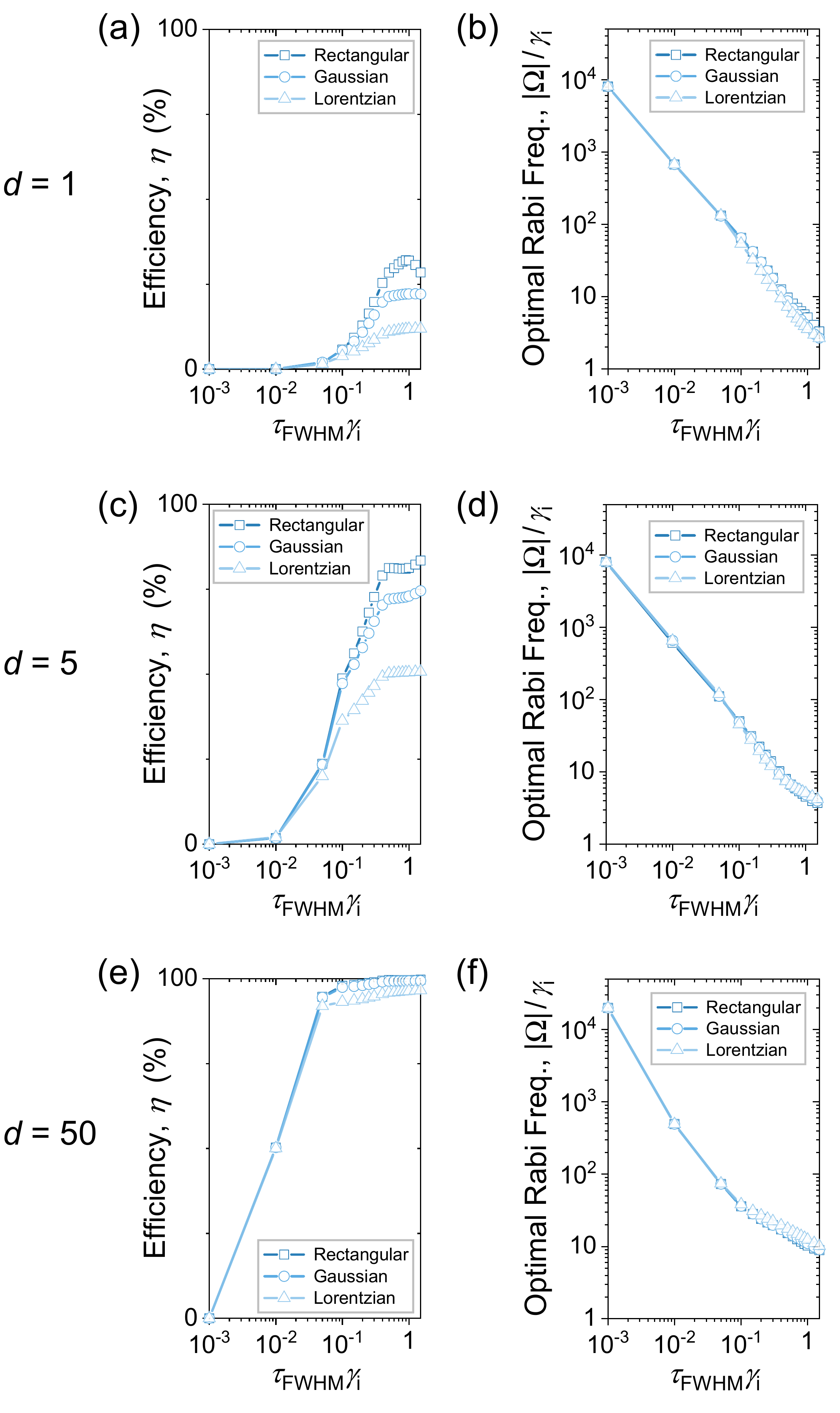}
	\caption{(a), (c), (e) Optimized memory efficiencies in inhomogeneous $\Lambda$-type level systems for Rectangular (square markers), Gaussian (circles), and Lorentzian (triangles) inhomogeneous distributions in the regime of intermediate optical depth and broadband pulse-duration--linewidth product, $\tFWHM\gamma_i$. At each point, memory efficiency is optimized with respect to the parameters of a Gaussian control field; the corresponding optimal control-field Rabi frequencies are shown in (b), (d), and (f), respectively.}
	\label{Fig_eta_opt}
\end{figure}

In Figure~\ref{Fig_eta_opt}, 
we present optimized inhomogeneous $\Lambda$-type memory efficiencies in the regime of intermediate optical depth ($d$) and pulse-duration--linewidth product ($\tFWHM\gamma_i$), for the same inhomogeneous lineshapes explored in Fig.~\ref{Fig_abs_dreXdw}. The large scale structure of each set of efficiencies is the same: near-unity efficiencies are possible for each inhomogeneous lineshape in the region of large optical depth and near matched signal bandwidth and ensemble linewidth (Fig.~\ref{Fig_eta_opt}(e); the region typically associated with the EIT protocol \cite{fleischhauer2000dark,shinbrough2021optimization,rastogi2019discerning}), and efficiencies tend toward 0 as $d\rightarrow0$ or 
when the signal bandwidth dramatically exceeds the inhomogeneous linewidth, $\tFWHM\gamma_i\ll 1$. The optimized control-field Rabi frequencies leading to these efficiencies are provided in Fig~\ref{Fig_eta_opt}(b), (d), and (f), demonstrating that each lineshape makes use of approximately the same control field power. The increased memory efficiency for the Rectangular distribution relative to the Gaussian and Lorentzian distributions, particularly in Fig.~\ref{Fig_eta_opt}(a) and (c), is therefore primarily due to the difference in inhomogeneous lineshape, not significant differences in control field optimization. 
The largest increase in memory efficiency we observe occurs for $d=5$, a typical optical depth for REI experiments, where the Rectangular distribution achieves roughly 32\% (8\%) enhancement in memory efficiency relative to the Lorentzian (Gaussian) profile. Qualitatively, these results agree with the simple model of Fig.~\ref{Fig_abs_dreXdw}, which predicts higher EIT efficiencies for the Rectangular profile. 

The intuition we develop for the presence of these relatively high efficiencies in the absence of complex inhomogeneous profile shaping can be separated into the ATT and EIT regimes: In the ATT regime, the signal field bandwidths are significantly larger than the inhomogeneous linewidth ($BW\gg\gamma_i$), and the $\pi$-pulse transfer of the frequency-multimode atomic polarization occurs significantly faster than the atomic polarization decoheres ($1/\Deltau \gg \gamma_i$). This ensures coherent population transfer, storage, and subsequent retrieval in the $\Lambda$-system. In the EIT regime, as we have seen, the destructive interference of many homogeneous EIT transparency windows leads to high-visibility inhomogeneous EIT, and this in turn leads to little atomic population entering the intermediate excited states during storage and retrieval operations. There is therefore little dephasing during storage and retrieval (i.e., adiabatic elimination of the intermediate excited states), which allows for high-efficiency memory operation.

\begin{figure}[t]
	\centering
	\includegraphics[width=0.9\columnwidth]{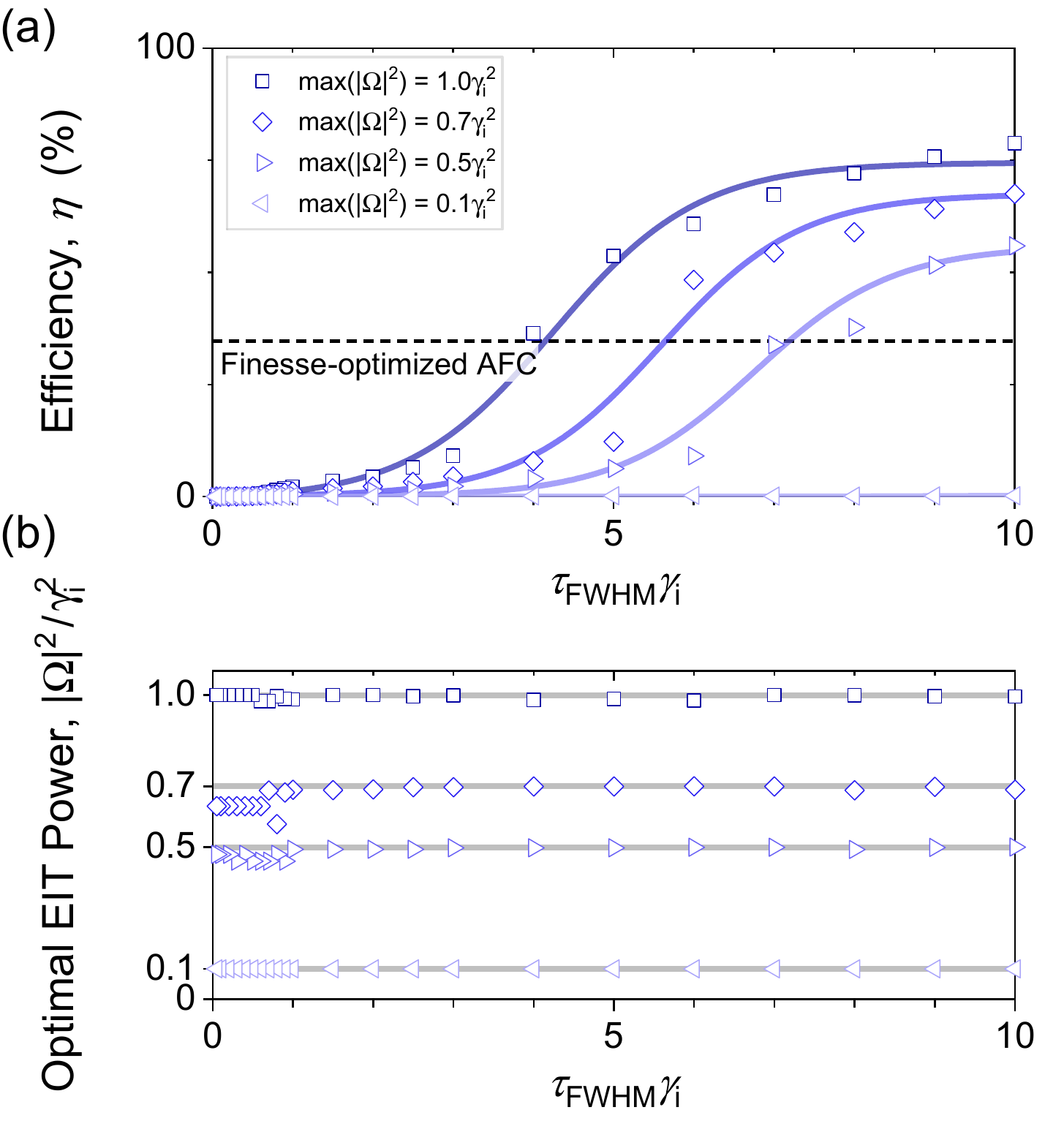}
	\caption{(a) Optimized EIT memory efficiencies in inhomogeneous $\Lambda$-type level systems assuming a Rectangular inhomogeneous distribution versus finesse-optimized AFC memory efficiencies, as a function of pulse-duration--linewidth product and maximum available control field power [markers: simulation data; solid lines: spline fits to guide the eye; dashed line: efficiency of finesse-optimized AFC that is an overestimate for $\tFWHM\gamma_i\lesssim1$  (see text)]. (b) Optimal EIT control field powers corresponding to the efficiencies in (a). Grey lines represent the maximum available control field power in each simulation.}
	\label{Fig_EIT_v_AFC}
\end{figure}

While near-unity efficiencies are possible using the EIT protocol in inhomogeneously broadened media, the optical depths required are on the threshold of what is typically possible in REI-doped solids in a multipass configuration \cite{riedel2022synthesis,shinbrough2023broadband}, and the control field powers required 
may be difficult to generate experimentally. In order to compare the EIT protocol in inhomogeneously broadened media with the AFC protocol in an experimentally realistic setting, we consider a conservative peak optical depth of $d=5$ and restrictions on the maximum available control field power, $\max(\abs{\Omega}^2)$. As is frequently the case in practice, this model ensures that the maximum achievable control field Rabi frequency is set by the available control field power. In Figure~\ref{Fig_EIT_v_AFC}, we present a comparison of the efficiencies achieved with the two protocols as a function of pulse-duration--linewidth product and varying maximum available control field power. The AFC protocol with optimal finesse $\mathcal{F}_\text{opt}$ and backward-propagating retrieval fields possesses an optimal memory efficiency given by $\eta_\text{opt}^\text{AFC} = (1-e^{-d_\text{eff}})^2 e^{-(1/\mathcal{F}_\text{opt}^2)\pi^2/2\ln2}$. This analytic expression applies for the case of narrowband signal fields, where $\tFWHM\gamma_i\gg 1$ (where, for the AFC protocol, $\gamma_i = \Gamma^\text{AFC}_i/2$); for a signal field that is broader than the comb, this efficiency 
is generally reduced due to poor spectral overlap of the signal field and AFC. In this example, trivial analytic optimization yields $\eta^\text{opt}_\text{AFC}=34.7\%$ for $\mathcal{F}_\text{opt} = 3.9$, as shown in Fig.~\ref{Fig_EIT_v_AFC}(a) (black, dashed line). For the EIT protocol, we again allow for optimization of the control field parameter space vector $\mathcal{G}$, with the constraint 
$\theta \leq \tctrl\sqrt{\max(\abs{\Omega}^2)/(8\pi \ln2)}$ with $\eta$ as the objective function. For large control field powers [darker traces in Fig.~\ref{Fig_EIT_v_AFC}(a)] and large $\tFWHM\gamma_i$, the efficiency of the EIT protocol surpasses the AFC protocol, with a crossover point of $\max(\abs{\Omega}^2)>4\gamma_i/\tFWHM$. Physically, this means that if one has access to a control field capable of creating an EIT transparency window greater than the photon bandwidth, the EIT protocol will lead to higher memory efficiency than using the same ensemble in an AFC configuration.  
In Fig.~\ref{Fig_EIT_v_AFC}(b), we provide the optimized EIT control field powers corresponding to the data in Fig.~\ref{Fig_EIT_v_AFC}(a); the optimized control field power saturates the maximum available power over the full range of each simulation, except for some small values of $\tFWHM\gamma_i$ where the EIT memory efficiency is low and the optimization algorithm can not find a sufficiently large efficiency gradient to arrive at the global optimum.

\section{Optimal Inhomogeneous Lineshape}\label{optSec}

In Sec.~\ref{surveySec} we examined a survey of common inhomogeneous lineshapes, and the resulting linear susceptibilities and quantum memory efficiencies. Here we turn to calculation of the theoretical optimal lineshape, for a given set of memory and control field parameters. The calculation proceeds as follows: We consider the linear integral transformation

\begin{equation}
    A_\text{out}(\tau) = \int d\tau'\, K(\tau,\tau') A_\text{in}(\tau')
\end{equation}

\noindent governing the combined storage and retrieval operations \cite{gorshkov2007photon_2,nunn2008quantum} for $\mathcal{M}_t=(5,1)$, which is near the onset of the EIT regime, with optimal control field parameters $\mathcal{G}_t=(2.75\pi,-0.25\tFWHM,1.25\tFWHM)$ for an initially Rectangular inhomogeneous lineshape. We construct $K(\tau,\tau')$ numerically by considering $A_\text{in}(\tau'_k) = \delta_D(\tau'_k)$, where $\delta_D(\tau)$ is the Dirac delta function, and integrating Eqs.~\eqref{Aeq}-\eqref{Beq} to find $A_\text{out}(\tau)=K(\tau,\tau'_k)$. We then take the singular value decomposition 

\begin{equation}
K(\tau,\tau') = \sum_l \lambda_l \psi_l(\tau)\phi^*_l(\tau')
\end{equation}

\noindent to find the optimal memory efficiency $\eta^\text{opt}=\max(\lambda_l^2)=\lambda_m^2$ and optimal input signal field envelope $A^\text{opt}_\text{in}(\tau)$~$=$~$\phi_m(\tau)$. We then construct $N=21$ interpolated functions $A^{(n)}_\text{in}(\tau)$ between $A^{(1)}_\text{in}(\tau) = A^\text{opt}_\text{in}(\tau)$ and the target Gaussian input signal field $A^{(N)}_\text{in}(\tau) = 
e^{-\tau^2/4\sigma^2}$. We fit 51 spline points to the (initially rectangular) inhomogeneous lineshape, and optimize over the vector of spline points $\mathcal{Z}$ using the same Nelder-Mead simplex method used in Sec.~\ref{surveySec} for an input signal field of $A^{(2)}_\text{in}(\tau)$. We use the resulting optimized spline points $\mathcal{Z}^{(2)}$ as an initial guess for the optimal inhomogeneous lineshape when $A^{(3)}_\text{in}(\tau)$ is used as an input signal field, and repeat this procedure iteratively until we find the optimal spline points $\mathcal{Z}^{(N)}$ defining the optimal inhomogeneous lineshape for $\mathcal{M}_t$ and $\mathcal{G}_t$. Due to the computational complexity of this task, we perform this optimization only for the selected $\mathcal{M}_t$ and $\mathcal{G}_t$, which are experimentally accessible and representative of the intermediate regime considered in this article, but this procedure can naturally be performed for additional $\mathcal{M}$ and $\mathcal{G}$, as desired.

\begin{figure}[t]
	\centering
	\includegraphics[width=1\columnwidth]{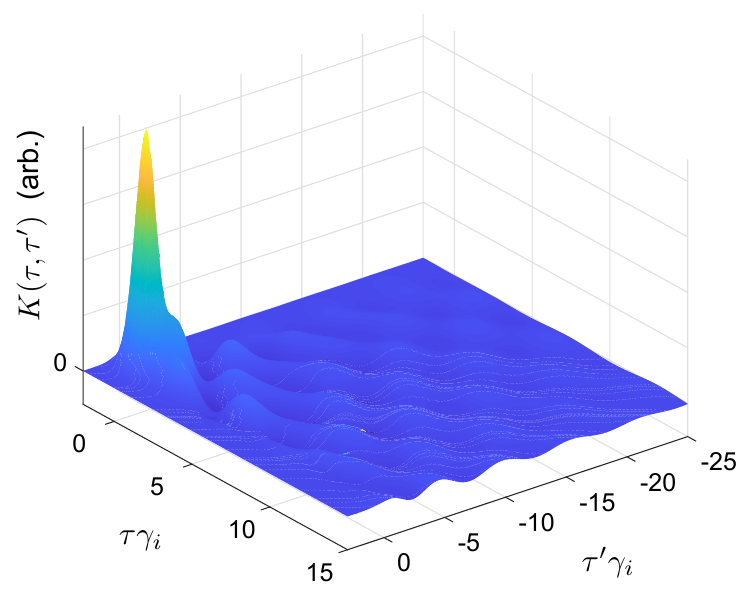}
	\caption{Numerically constructed linear integral kernel $K(\tau,\tau')$ governing storage and retrieval in an inhomogeneously broadened ensemble of $\Lambda$-type emitters with memory parameters $\mathcal{M}_t = (5,1)$ and control field parameters $\mathcal{G}_t=(2.75\pi,-0.25\tFWHM,1.25\tFWHM)$. }
	\label{Fig_kernel}
\end{figure}

Figure~\ref{Fig_kernel} shows a region of interest of the integral kernel $K(\tau, \tau')$ we numerically construct for the target $\mathcal{M}_t$ and $\mathcal{G}_t$ (the full integral kernel used in the following calculations is evaluated over the range $\tau'\gamma_i\in[-60,+10]$ and $\tau\gamma_i\in[-5,+70]$, which ensures sampling of $K(\tau, \tau')$ to $<10^{-5}$ of its maximal value). We follow the procedure described above to determine the optimal inhomogeneous profile $\mathcal{Z}^{(N)}$ for the chosen memory and control field parameters, and we find no change in the optimal inhomogeneous lineshape from the initial Rectangular distribution within a numerical accuracy of $\sim0.5\%$ in memory efficiency. This result implies that the Rectangular distribution is an extremal case, and represents the optimal inhomogeneous distribution for EIT-like quantum memory in inhomogeneously broadened ensembles. To confirm this result, we repeat the interpolation and optimization procedure for 100 randomly distributed initial inhomogeneous profiles, as well as several hand-selected profiles close but not identical to a Rectangular distribution (i.e., a Rectangular distribution modified by a quadratic component $\pm0.1\delta^2$ to peak or round the cusps of the Rectangular distribution); in each case, the optimal $\mathcal{Z}^{(N)}$ returns to a Rectangular distribution within numerical accuracy.

\section{Conclusion}

We have shown the potential for improving quantum light storage by unconstrained tailoring of the spectral profile of an inhomogeneously broadened atomic ensemble. We see that a highly irregular rectangular shape enables substantially more efficient storage for low to moderate optical depths without the need for additional control field power. We also show that the popular atomic frequency comb protocol, which uses spectral tailoring of such an ensemble to store light, is not an optimized use of the available atoms when control field power is unrestricted or when the signal field is narrowband relative to the inhomogeneous distribution. Finally, we present a method for calculating the optimal inhomogeneous lineshape, which we apply for a particular set of parameters, revealing the optimality of the rectangular distribution. Natural extensions of this work include inhomogeneous lineshape
optimization in other parameter regimes (e.g., considering larger optical depths), accounting for additional experimental constraints, including protocols beyond the EIT-type storage considered here, and including inhomogeneity in the ground to metastable state transition. This work and these generalizations greatly expand the toolbox for efficient, broadband, and long-lived quantum memory operation in inhomogeneously broadened ensembles of quantum emitters.


\ \\

\section*{Acknowledgements}

We gratefully acknowledge helpful discussion provided by Yujie Zhang, Kathleen Oolman, Dongbeom Kim, Ashwith Prabhu, and Priyash Barya. This work was supported by NSF Grant No.~2207822 and by the U.S. Department of Energy, Office of Science, National Quantum Information Science Research Centers. This work made use of the Illinois Campus Cluster, a computing resource that is operated by the Illinois Campus
Cluster Program (ICCP) in conjunction with the National Center for Supercomputing Applications (NCSA) and which is supported by funds from the University of Illinois Urbana-Champaign.

\appendix

\renewcommand\thefigure{\thesection.\arabic{figure}}

\setcounter{figure}{0} 



%

\end{document}